\documentclass[12pt]{article}

\newcommand{\nc}{\newcommand}
\nc{\gtwid}{\mathrel{\raise.3ex\hbox{$>$\kern-.75em\lower1ex\hbox{$\sim$}}}}
\nc{\ltwid}{\mathrel{\raise.3ex\hbox{$<$\kern-.75em\lower1ex\hbox{$\sim$}}}}
\nc{\comp}{{\rm C}\llap{\vrule height7.1pt width1pt depth-.4pt\phantom t}}

\usepackage{pqft}
\usepackage{epsfig}

\begin{document}

\title{Fermion Self-Energy during Inflation}

\author{R. P. Woodard}

\institutes{Department of Physics \\ University of Florida \\
Gainesville, FL 32611 \\ UNITED STATES} 

\maketitle\abstract{I report on work done with Tomislav Prokopec
(Heidelberg, CERN). We computed the one loop self-energy of a
massless fermion during inflation. When the fermion is free it
experiences only a small amount of inflationary particle production
owing to the conformal invariance of its classical action. However, 
when the fermion is Yukawa coupled to a massless, minimally coupled 
scalar, there is copious production of fermions. In a more 
complicated model this effect might generate baryon asymmetry 
during inflation.}

\authorrunning{R. P. Woodard}

\titlerunning{Fermion Self-Energy during Inflation}

\section{Introduction}

\begin{figure}[htbp]
\vskip 0.1in
\leftline{\hskip 0.1in\epsfig{file=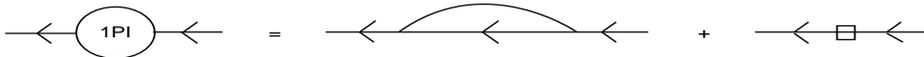,
                               width=7.0in,height=7.0in }}
\vskip -6.5in
\caption{ {\it The one loop fermion self-energy.} Fermion lines have an
arrow, scalar lines do not. The final diagram gives the contribution of 
field strength renormalization. \label{fig1} }
\end{figure}

Tomislav Prokopec and I have recently computed the one loop self-energy,
during de Sitter inflation, of a massless fermion which is Yukawa coupled
to a massless, minimally coupled scalar. The diagrammatic representation of
what we did is given in Fig.~\ref{fig1}. This would be a trivial exercise in 
flat space, and a fairly uninteresting one. The inflationary background 
geometry makes the calculation both nontrivial and physically interesting. 
What we found is that inflation engenders copious production of scalars
which then decay into fermion-anti-fermion pairs. The first fact was known
but the second is novel and potentially quite significant. In a more 
complicated theory it may provide a mechanism for generating a baryon
asymmetry {\it during} inflation.

Most of the theoretical technology we used was developed in a previous 
computation, with Ola T\"ornkvist, of the one loop vacuum polariztion 
induced by massless scalar QED during inflation \cite{PRL,AOP}. However, 
the result is very different. Whereas the photon develops a mass, which
highly {\it suppresses} the production of photons during inflation, we
find copious production of Yukawa coupled fermions. I will therefore devote 
most of this talk to explaining the physics in very simple terms. In Section 
2 I give a brief review of inflation. The largest part of the talk is 
Section 3. In it I explain why the direct production of massless, minimally 
coupled scalars occurs during inflation, and why there is no significant 
direct production of fermions. Section 4 presents the result, without 
giving the derivation, and shows how it implies fermion production.

\section{Inflation}

On the largest scales the universe is amazingly homogeneous and isotropic
\cite{KT}. It also seems to be devoid of spatial curvature \cite{WMAP}.
The spacetime geometry consistent with these three features is characterized 
by the following simple invariant element,
\begin{equation}
ds^2 = -dt^2 + a^2(t) d\vec{x} \cdot d\vec{x} \; . \label{ds^2}
\end{equation}
The coordinate $t$ represents physical time, the same as it does in flat 
space. However, the physical distance between $\vec{x}$ and $\vec{y}$ is
not given by their Euclidean norm, $\Vert \vec{x} - \vec{y} \Vert$, but
rather by $a(t) \Vert \vec{x} - \vec{y} \Vert$. Because it converts coordinate
distance into physical distance $a(t)$ is known as the {\it scale factor}.

Although the scale factor is not directly measurable, three simple observable
quantities can be constructed from it,
\begin{equation}
z \equiv \frac{a_0}{a(t)} - 1 \quad , \quad 
H(t) \equiv \frac{\dot{a}}{a} \quad , \quad q(t) \equiv - 
\frac{a \ddot{a}}{\dot{a}^2} = -1 - \frac{\dot{H}}{H^2} \; .
\end{equation}
The {\it redshift} $z$ gives the proportional increase in the wavelength of 
light emitted at time $t$ and received at the current time, $t_0$. Redshift 
is often used to measure cosmological time, even for epochs from which we 
detect no radiation. The {\it Hubble parameter} $H(t)$ gives the rate at 
which the universe is expanding. It's current value is, $H_0 = (71 {+4 
\atop -3}) \frac{\rm km}{\rm s\ Mpc} \simeq 2.3 \times 10^{-18}~{\rm Hz}$ 
\cite{WMAP}. The {\it deceleration parameter} $q(t)$ is less well 
measured. Observations of Type Ia supernova are consistent with a current 
value of $q_0 \simeq -.6$ \cite{SN}.

{\it Inflation} is defined as a phase of accelerated expansion, that is, 
$q(t) < 0$ with $H(t) > 0$. From the current values of the cosmological
parameters one can see that the universe seems to be in such a phase now. 
However, I wish to discuss {\it primordial inflation}, which is conjectured 
to have occurred at something like $10^{-37}$ seconds after the beginning 
of the universe with a Hubble parameter 55 orders of magnitude larger than 
it is today. There are many reasons for believing that the very early 
universe underwent such a phase \cite{Linde}. I will confine myself to 
reviewing how inflation resolves the {\it smoothness problem}. This can be 
summed up in the question, why does the large scale universe possess such a 
simple geometry (\ref{ds^2})?

To understand the problem we need to compare the distance light can travel
from the beginning of the universe to the time of some observable event, with 
the distance it can travel from then to the present. From the invariant 
element (\ref{ds^2}) we see that the radial position of a light ray obeys, 
$dr = \pm dt/a(t)$. The minus sign gives the {\it past light-cone} of the
point $x^{\mu} = (t_0,\vec{0})$, whereas the plus sign gives the {\it future 
light-cone} of a point $x^{\mu} = (t_i,\vec{0})$ at the beginning of the 
universe,
\begin{equation}
R_{\rm past} = \int_{t_{obs}}^{t_0} \frac{dt}{a(t)} \qquad , \qquad 
R_{\rm future} = \int_{t_i}^{t_{obs}} \frac{dt}{a(t)} \; . \label{lcones}
\end{equation}
We can observe thermal radiation from the time of decoupling ($z_{dec} 
\simeq 1089$) whose temperature is isotropic to one part in $10^5$. Unless 
the universe simply began this way --- which seems unlikely --- 
equilibrium must have been established by causal processes. In other words, 
we must have $R_{\rm future} > R_{\rm past}$.

Suppose that, during the period $t_1 \le t \le t_2$, the deceleration 
parameter is constant $q(t) = q_1$. In that case we can obtain explicit 
expressions for the Hubble parameter and the scale factor in terms of their 
values at $t=t_1$,
\begin{equation}
H(t) = \frac{H_1}{1 + (1 \!+\! q_1) H_1 (t \!-\! t_1)} \quad {\rm and } 
\quad a(t) = a_1 \Bigl[1 + (1 \!+\! q_1) H_1 (t \!-\! t_1)\Bigr]^{\frac1{1 
\!+\!q_1}} \; .
\end{equation}
These expressions permit us to evaluate the fundamental integral involved
in the past and future light-cones (\ref{lcones}),
\begin{equation}
\int_{t_1}^{t_2} \frac{dt}{a(t)} = \frac1{a_1 H_1 q_1} \Bigl[1 \!+\!
(1\!+\! q_1) H_1 (t \!-\! t_1)\Bigr]^{\frac{q_1}{1\!+\!q_1}}
\Bigl\vert_{t_1}^{t_2} =  \frac1{q_1} \left\{ \frac1{a_2 H_2} - 
\frac1{a_1 H_1} \right\} .
\end{equation}

Although $q_0$ is negative, this is a recent event ($z \simeq 1$) which 
followed a long period of nearly perfect matter domination with $q = 
+\frac12$. Much before the time of matter-radiation equality ($z_{eq} 
\simeq 3200$) the universe was almost perfectly radiation-dominated, which 
corresponds to $q = +1$. To simplify the computation we will ignore the
the recent phase of acceleration and also the transition periods,
\begin{equation}
a(t) H(t) = a_0 H_0 \cases{ \sqrt{1 + z} & $\forall \, z \le z_{eq}$ \cr
\frac{1 + z}{\sqrt{1 + z_{eq}}} & $\forall \, z \ge z_{eq}$} \; .
\label{ass}
\end{equation}
The cosmic microwave radiation was emitted within about a hundred redshifts 
of $z_{dec} < z_{eq}$, so the past light-cone is,
\begin{equation}
R_{\rm past} = \frac2{a_0 H_0} \left\{ \frac1{\sqrt{1 \!+\! 0}} - \frac1{
\sqrt{1 \!+\! z_{dec}}} \right\} \simeq \frac2{a_0 H_0} \; . \label{past}
\end{equation}
The future light-cone derives from both epochs and it depends slightly upon 
the beginning redshift, $z_{beg}$,
\begin{equation}
R_{\rm future} = \frac2{a_0 H_0} \left\{ \frac1{\sqrt{1 \!+\! z_{dec}}} - 
\frac1{\sqrt{1 \!+\! z_{eq}}} \right\} + \frac1{a_0 H_0} \left\{\frac{\sqrt{1 
\!+\! z_{eq}}}{1 \!+\! z_{eq}} - \frac{\sqrt{1 \!+\! z_{eq}}}{1 \!+\! z_{beg}}
\right\} \; .
\end{equation}
One maximizes $R_{\rm future}$ by taking $z_{beg} \rightarrow \infty$, but it
isn't enough. Under the assumption of $q = +1$ before $z_{eq}$ we are forced
to conclude that the 2-dimensional surface we can see from the time of 
decoupling consists of $(R_{\rm past}/R_{\rm future})^2 \simeq 2200$ 
regions which cannot have exchanged even a photon since the beginning of 
time! So how did they reach equilibrium?

This embarrassment resulted from the fact that the upper limit of integration 
dominates $R_{\rm future}$ for positive deceleration. Inflation solves the
problem by positing a very early epoch of negative deceleration. This makes 
the lower limit of $R_{\rm future}$ dominate, so the future light-cone can be 
made as large as necessary by increasing $z_{beg}$

\section{Inflationary Particle Production}

\subsection{Virtual particles in flat space}

To understand why certain kinds of virtual particles can be ripped out of 
the vacuum during inflation it is instructive to review what constrains the
lifetimes of virtual particles in flat space. A particle with wave vector 
$\vec{k}$ and mass $m$ has energy,
\begin{equation}
E(\vec{k}) = \sqrt{m^2 + \Vert \vec{k} \Vert^2} \; .
\end{equation}
According to quantum field theory, virtual particles are continually
emerging from the vacuum with all different momenta. This process can 
conserve momentum if a virtual pair emerges with opposite 3-momenta, but
it cannot conserve energy. Before the pair came into existence the energy
was zero, afterwards it was $2 E(\vec{k})$. The energy-time uncertainty
principle asserts that we cannot detect this violation provided the pair 
annihilates in less than a time ${\Delta t}$ given by the inequality,
\begin{equation}
\int_t^{t + \Delta t} \!\!\!\!\!\! dt' \, 2 E(\vec{k}) \ltwid 1 \qquad 
\Longrightarrow \qquad \Delta t \ltwid \frac1{2 E(\vec{k})} \; .
\end{equation}

Of course the formalism of quantum field theory automatically incorporates 
these effects, without the need to invoke additional principles. However, 
the energy-time uncertainty principle allows one to understand many things. 
For example, virtual electron-positron pairs polarize the vacuum the most 
because they are the lightest charged particles and hence have the longest 
time to align themselves with applied fields. Similarly, it is long wave 
length virtual particles that survive the longest, which is one way of
understanding long range forces.

\subsection{Virtual particles in cosmology}

In a homogeneous and isotropic geometry (\ref{ds^2}) particles are still
labeled by their constant, comoving wave vectors. However, because $\vec{k}$
involves an inverse wave length it must be divided by the scale factor in
computing physical quantities such as the energy,
\begin{equation}
E(t,\vec{k}) = \sqrt{m^2 + \Vert \vec{k} \Vert^2/a^2(t)} \; .
\end{equation}
Let us now reconsider the emergence of a virtual pair of such particles
with wave vectors $\pm \vec{k}$. If they emerge at time $t$, the energy-time
uncertainty principle implies we will detect no violation of energy
conservation provided they annihilate within a time $\Delta t$ such that,
\begin{equation}
\int_t^{t + \Delta t} \!\!\!\!\!\! dt' \, 2 E(t',\vec{k}) \ltwid 1 \; .
\label{unc}
\end{equation}

What can increase $\Delta t$? Obviously anything that reduces $E(t',\vec{k})$.
At fixed wave vector $\vec{k}$ and time $t'$ this is accomplished by taking 
the mass to zero. Zero mass simplifies the integrand in (\ref{unc}) to $2 \Vert 
\vec{k} \Vert/a(t)$. Up to a constant this is the same expression as in the
light-cones (\ref{lcones})! We have already seen that negative deceleration
(and hence inflation) causes these integrals to be dominated by their lower
limits. Hence they cannot become arbitrary large as $\Delta t$ goes to
infinity. The most negative deceleration consistent with stability is $q_i = 
-1$, for which the uncertainty bound gives,
\begin{equation}
m = 0 \; {\rm and} \; a(t) = a_i e^{H_i t} \quad \Longrightarrow \quad
\int_t^{t + \Delta t} \!\!\!\!\!\! dt' \, 2 E(t,\vec{k}) = \frac{2 \Vert 
\vec{k} \Vert}{H_i a(t)} \Bigl[1 - e^{-H_i \Delta t}\Bigr] \; . \label{why}
\end{equation}
We therefore conclude that any virtual particle pair which emerges with $\Vert 
\vec{k} \Vert \ltwid H_i a(t)$ can survive for ever.

\subsection{Conformal invariance is the kiss of death}

One might imagine that the big obstacle to inflationary particle production
is nonzero mass, but this is not so. At the enormous energy scales envisaged
for primordial inflation all the known particles are effectively massless.
The reason most of them still do not experience inflationary particle 
production is that they possess a symmetry suppressing the rate at which 
virtual particles emerge from the vacuum.

The killer symmetry is known as {\it conformal invariance}. It rescales the 
fields which carry scalars ($\phi$), spin $\frac12$ fermions ($\psi$), 
vectors ($A_{\mu}$) and gravitons ($g_{\mu\nu}$) by powers of an arbitrary 
function of space and time $\Omega(x)$,
\begin{eqnarray}
\phi(x) \rightarrow \Omega^{\!-\!1}\!(x) \phi(x) & , & A_{\mu}(x) \rightarrow
A_{\mu}(x) \; , \\
\psi(x) \rightarrow \Omega^{\!-\!\frac32}\!(x) \psi(x) & , & g_{\mu\nu}(x) 
\rightarrow \Omega^2\!(x) g_{\mu\nu}(x) \; .
\end{eqnarray}
A simple, conformally invariant theory is electromagnetism,
\begin{equation}
{\cal L}_{\rm EM} = -\frac14 F_{\alpha\beta} F_{\rho \sigma} g^{\alpha \rho} 
g^{\beta \sigma} \sqrt{-g} \; ,
\end{equation}
where $F_{\mu\nu} \equiv \partial_{\mu} A_{\nu} - \partial_{\nu} A_{\mu}$.
Since the vector potential is unaffected by a conformal transformation, the
Lagrangian's invariance follows from the combination of inverse metrics and
the square root of the determinant,
\begin{equation}
g^{\alpha \rho} g^{\beta \sigma} \sqrt{-g} \rightarrow \Omega^{-2} g^{\alpha
\rho} \cdot \Omega^{-2} g^{\beta \sigma} \cdot \Omega^4 \sqrt{-g} =
g^{\alpha \rho} g^{\beta \sigma} \sqrt{-g} \; .
\end{equation}

The connection between conformal invariance and cosmology derives from
the existence of a coordinate system in which the general homogeneous and 
isotropic metric (\ref{ds^2}) becomes just $g_{\mu\nu} = a^2 \eta_{\mu\nu}$,
\begin{equation}
dt = a d\eta \qquad \Longrightarrow \qquad ds^2 = a^2 \Bigl(-d\eta^2 +
d\vec{x} \cdot d\vec{x} \Bigr) = a^2 \eta_{\mu\nu} dx^{\mu} dx^{\nu} \; .
\end{equation}
If we take a conformally invariant Lagrangian and express it in terms of 
the conformally rescaled fields with $\Omega = 1/a$, it is just as if the
theory was in flat space! Some examples are the conformally coupled scalar,
massless Dirac fermions, and electromagnetism,
\begin{eqnarray}
{\cal L}_{\rm CCS} = -\frac12 \partial_{\mu} \phi \partial_{\nu} \phi 
g^{\mu\nu} \sqrt{-g} - \frac1{12} \phi^2 R \sqrt{-g} & = & -\frac12 
\partial_{\mu} (a \phi) \partial_{\nu} (a \phi) \eta^{\mu\nu} \; , \\
{\cal L}_{\rm Dirac} = \overline{\psi} e^{\mu}_{~b} \gamma^b \Bigl( i
\partial_{\mu} - \frac12 A_{\mu c d} J^{cd}\Bigr) \psi \sqrt{-g} & = & 
(a^{\frac32} \overline{\psi}) \gamma^{\mu} i \partial_{\mu} (a^{\frac32} \psi)
\; , \label{Dirac} \\
{\cal L}_{\rm EM} = -\frac14 F_{\alpha \beta} F_{\rho \sigma} g^{\alpha \rho} 
g^{\beta \sigma} \sqrt{-g} & = & -\frac14 F_{\alpha \beta} F_{\rho \sigma}
\eta^{\alpha \rho} \eta^{\beta \sigma} \; .
\end{eqnarray}

Physics does not depend upon local, invertible field redefinitions, so
a conformally invariant theory cannot be different, in conformal coordinates,
than it is in flat space. This applies to all physical quantities, including 
the rate at which virtual particles emerge from the vacuum, per unit conformal
time. Converting back to physical time gives the stated suppression,
\begin{equation}
\frac{dn}{dt} = \frac{d\eta}{dt} \frac{dn}{d\eta} = \frac1{a} \, 
\Bigl({\rm flat\ space\ rate}\Bigr) \; .
\end{equation}
Therefore any conformally invariant, massless virtual particles which happen 
to emerge with $\Vert \vec{k} \Vert \ltwid H_i a(t)$ will become real, but
not many will emerge.

\subsection{Massless minimally coupled scalars}

Most familiar particles become (classically) conformally invariant when their
masses are taken to zero. The exceptions are gravitons and massless, minimally
coupled scalars. The Lagrangian density for the later is,
\begin{equation}
{\cal L}_{\rm MMCS} = -\frac12 \partial_{\mu} \phi \partial_{\nu} \phi 
g^{\mu \nu} \sqrt{-g} = -\frac12 a^2 \partial_{\mu} \phi \partial_{\nu}
\phi \eta^{\mu\nu} \; .
\end{equation}
Integrating over space to obtain the Lagrangian, then using Parseval's
theorem to convert to Fourier space, gives a form we can recognize,
\begin{equation}
L_{\rm MMCS} \equiv \int d^3x {\cal L}_{\rm MMCS} = \frac12 \int 
\frac{d^3k}{(2 \pi)^3} \Bigl\{ a^3(t) \vert \dot{\widetilde{\phi}}(t,\vec{k})
\vert^2 - a(t) \Vert \vec{k} \Vert^2 \vert \widetilde{\phi}(t,\vec{k}) \vert^2 
\Bigr\} .
\end{equation}
Each wave vector $\vec{k}$ represents an independent harmonic oscillator
with a time dependent mass, $m(t) \sim a^3(t)$, and frequency, $\omega(t)
= \Vert \vec{k} \Vert/a(t)$!

For the case of de Sitter inflation ($a(t) = a_i e^{H_i t}$) the mode
functions take a simple form,
\begin{equation}
\widetilde{\phi}(t,\vec{k}) = u(t,k) a(\vec{k}) + u^*(t,k) 
a^{\dagger}(-\vec{k}) \; {\rm where} \; u(t,k) \equiv \frac{H_i}{
\sqrt{2 k^3}} \Bigl(1 - \frac{i k}{a H_i} \Bigr) e^{\frac{i k}{a H_i}} .
\end{equation}
To understand the meaning of the operators $a(\vec{k})$ and $a^{\dagger}(
\vec{k})$, note that the minimum energy in wave vector $\vec{k}$ at time $t$ is 
$\omega(t)$. However, the state with this energy does not evolve onto itself. 
Indeed, this theory has no stationary states! A reasonable ``vacuum'' is the 
state that was minimum energy in the distant past. This is known as {\it 
Bunch-Davies vacuum} and it is defined by $a(\vec{k}) \vert \Omega \rangle = 
0$.

For Bunch-Davies vacuum the 0-point energy in wave vector $\vec{k}$ is,
\begin{equation}
E_0(t,\vec{k}) = \frac12 a^3(t) \vert \dot{u}(t,k) \vert^2 + \frac12 a(t)
\Vert \vec{k} \Vert^2 \vert u(t,k) \vert^2 = \frac{\Vert \vec{k} \Vert}{2 a(t)}
+ \frac{H_i^2 a(t)}{4 \Vert \vec{k} \Vert} \; .
\end{equation}
The first term on the far right represents the irreducible, minimum energy.
The second term gives the extra energy due to inflationary particle production.
Since the energy of a particle of wave number $\vec{k}$ is $\Vert \vec{k}
\Vert/a(t)$ we can easily compute the number of particles,
\begin{equation}
N(t,\vec{k}) = \left( \frac{H_i a(t) }{2 \Vert \vec{k} \Vert } \right)^2 \; .
\end{equation}
As one might expect from the preceding discussion, this is much less than
one at very early times, and it becomes order one when the wave number just
begins to satisfy the uncertainty bound (\ref{why}).

\subsection{Scalar decay to fermions}

Although we have seen that inflation produces enormous numbers of massless,
minimally coupled scalars, the conformal invariance of the Dirac Lagrangian
(\ref{Dirac}) implies that there can be no comparable, direct production of 
fermi\-ons. However, it is still possible to make lots of fermions during
inflation by allowing the scalars to decay into them. This can be 
accomplished by making scalars and fermions interact through a Yukawa 
coupling,
\begin{eqnarray}
{\cal L}_{\rm Yukawa} & = & {\cal L}_{\rm MMCS} + {\cal L}_{\rm Dirac} -
\Gamma \phi \overline{\psi} \psi \sqrt{-g} , \qquad \\
& = & -\frac12 a^2 \partial_{\mu} \phi \partial_{\nu} \phi 
\eta^{\mu\nu} \!+\! (a^{\frac32} \overline{\psi}) \gamma^{\mu} i \partial_{\mu}
(a^{\frac32} \psi) \!-\! \Gamma a^4 \phi \overline{\psi} \psi . \label{Yuk}
\end{eqnarray}

\begin{figure}[htbp]
\vskip 0.1in
\leftline{\hskip 1.5in\epsfig{file=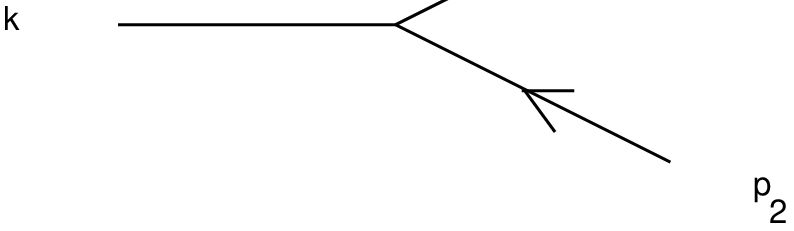,
                               width=7.0in,height=7.0in }}
\vskip -6.0in
\caption{ {\it Scalar decay into a fermion-anti-fermion pair.} The initial
state scalar has wave vector $\vec{k}$. The final state fermions have wave
vectors $\vec{p}_1 = \vec{p}$ and $\vec{p}_2 = \vec{k} - \vec{p}$.
\label{fig2} }
\end{figure}

The decay process is depicted in Fig.~\ref{fig2}. Just as with inflationary 
particle production, 3-momentum is conserved but energy is generally not,
\begin{equation}
\vec{k} = \vec{p} + (\vec{k} - \vec{p}) \qquad \Longrightarrow \qquad 
\Delta E(t) = \Bigl[\Vert \vec{p} \Vert + \Vert \vec{k} - \vec{p} \Vert - 
\Vert \vec{k} \Vert \Bigr] \frac1{a(t)} \; .
\end{equation}
In flat space the decay would be only be allowed for the infinitesimal phase
space in which three momenta are co-linear. However, just as for particle 
production, the inflationary expansion weakens the constraint imposed by
the energy-time uncertainty principle so as to permit the decay over a
large volume of phase space,
\begin{equation}
\int_t^{t+\Delta t} dt' \Delta E(t') \ltwid 1 \qquad \Longrightarrow \qquad
\Bigl[\Vert \vec{p} \Vert + \Vert \vec{k} - \vec{p} \Vert - \Vert \vec{k} 
\Vert \Bigr] \ltwid H a(t) \; .
\end{equation}
This is the physics behind our result.

\section{The Calculation}

I now specialize to the de Sitter scale factor, $a(t) = e^{H t} = -1/H\eta$.
It is useful to express the propagators in terms of the following conformal 
coordinate interval,
\begin{equation}
{\Delta x}^2(x;x') \equiv \Vert \vec{x} \!-\! \vec{x}' \Vert^2 - (\vert
\eta \!-\! \eta'\vert - i \epsilon)^2 \; . \label{Dx^2}
\end{equation}
Owing to the conformal invariance of the free Lagrangian (\ref{Dirac}), the 
fermion propagator is a trivial rescaling of the flat space result,
\begin{equation}
i S_{ij}(x;x') = (a a')^{\!-\frac32} \gamma^{\mu}_{ij} \, i \partial_{\mu} 
\left\{ \frac1{4 \pi^2} \frac1{{\Delta x}^2(x;x')} \right\} \; , \label{Sij}
\end{equation}
where $a \equiv a(t)$ and $a' \equiv a(t')$ are the scale factors evaluated
at the two points. The free scalar is not conformally invariant so its 
propagator has an additional term,
\begin{equation}
i \Delta(x;x') = \frac1{4 \pi^2} \left\{\frac{(a a')^{\!-1}}{
{\Delta x}^2(x;x')} - \frac1{2} H^2 \ln\Bigl[H^2 {\Delta x}^2(x;x')\Bigr]
\right\} \; . \label{Delta}
\end{equation}
The $\phi \overline{\psi}_i \psi_j$ vertex and the fermion field strength
renormalization are,
\begin{equation}
-i \Gamma a^4 \delta_{ij} \qquad {\rm and} \qquad i {\delta Z}_2 
(a a')^{\frac32} \gamma^{\mu}_{ij} \, i \partial_{\mu} \delta^4(x-x') \; .
\label{verts}
\end{equation}

Computing one loop diagrams in position space is easy. One simply multiplies
the various vertices and propagators. There are no integrations. The two
diagrams of Fig.~\ref{fig1} give,
\begin{eqnarray}
-i \Bigl[ \hbox{}_i \Sigma_j\Bigr](x;x') & = & \Bigl( -i \Gamma a^4 
\delta_{ik} \Bigr) i S_{k\ell}(x;x') \Bigl( -i \Gamma a^{\prime 4} 
\delta_{\ell j} \Bigr) i \Delta(x;x')  \nonumber \\
& & \hspace{2cm} + i {\delta Z}_2 (a a')^{\frac32} \gamma^{\mu}_{ij} \, i 
\partial_{\mu} \delta^4(x-x') \; . \label{1loop}
\end{eqnarray}
Well, it isn't {\it quite} that easy! The actual calculation was done
using the Schwinger-Keldysh formalism, which involves summing diagrams with
slight variations in the $i \epsilon$ prescription of expression (\ref{Dx^2})
\cite{Schwinger,Jordan}.

One must also account for the ultraviolet divergences which manifest as 
products that are not integrable functions of $x^{\mu}$ and $x^{\prime\mu}$.
One deals with these by partially integrating to reduce negative powers of
${\Delta x}^2(x;x')$ until the result is integrable. This procedure can be 
implemented so as to segregate the divergence to a delta function that can 
then be absorbed by $\delta Z_2$. Doing this rigorously requires the use of 
a regulator. We employed dimensional regularization, which makes slight 
changes in the fermion propagator (\ref{Sij}) and the vertices (\ref{verts}), 
and major changes in the scalar propagator (\ref{Delta}) \cite{phi^4}. 

Most of the formalism has been discussed in print \cite{AOP} so I will 
confine myself to quoting the fully renormalized result,
\begin{eqnarray}
\lefteqn{\Bigl[ \hbox{}_i \Sigma_j\Bigr](x;x') = \frac{-\Gamma^2}{2^8 \pi^3}
(a a')^{\frac32} \gamma^{\mu}_{ij} \, i \partial_{\mu} \partial^4 \left\{ 
\theta({\Delta \eta}) \theta({\Delta \eta} \!-\! {\Delta x}) \Bigl( \ln\Bigl[ 
\mu^2 ({\Delta \eta}^2 \!\!-\! {\Delta x}^2)\Bigr] \!-\! 1 \Bigr)\! \right\} } 
\nonumber \\
& & \hspace{3cm} - \frac{\Gamma^2}{2^5 \pi^2} \ln(a a') (a a')^{\frac32} 
\gamma^{\mu}_{ij} \, i \partial_{\mu} \delta^4(x-x') \nonumber \\
& & \hspace{1cm} - \frac{\Gamma^2 H^2}{2^6 \pi^3} (a a')^{\frac52} 
\gamma^{\mu}_{ij} \, i \partial_{\mu} \partial^2 \left\{\theta({\Delta \eta}) 
\theta({\Delta \eta} \!-\! {\Delta x}) \ln\Bigl[H^2 ({\Delta \eta}^2 \!-\! 
{\Delta x}^2)\Bigr] \right\} . \qquad \label{Sigma}
\end{eqnarray}
The first line is the conformally rescaled flat space result. Note that it
involves a renormalization scale $\mu$. The second line is the well-known 
contribution from the conformal anomaly. The intrinsic de Sitter result is on 
the third line. It is distinguished from the other two terms by its extra 
factor of $a a'$. 

The self-energy gives quantum corrections to the Dirac equation,
\begin{equation}
a^{\frac32} \gamma^{\mu}_{ij} i \partial_{\mu} \Bigl(a^{\frac32} \psi_i(x) 
\Bigr) + \int d^4x' \Bigl[ \hbox{}_i \Sigma_j\Bigr](x;x') \psi_j(x') = 0 \; .
\end{equation}
Note that the flat space part of the one loop self-energy (\ref{Sigma}) has 
the same number of scale factors as the tree order differential operator. 
Perturbation theory only makes sense if the coupling constant is small, so we 
may assume $\Gamma^2 \ll 1$. This means that the flat space part of the one 
loop self-energy can be ignored. So too can the conformal anomaly, which is 
only enhanced by a factor of $\ln(a)$. However, we cannot necessarily ignore 
the de Sitter correction. Although it is also down by $\Gamma^2$, it is 
enhanced by a scale factor, which becomes enormously large during inflation.
It therefore makes sense to keep only the de Sitter contribution of 
(\ref{Sigma}).

Because the background is spatially translation invariant it also makes sense
to look for plane wave solutions,
\begin{equation}
\psi_i(\eta,\vec{x}) = a^{-\frac32} \chi_i(\eta) e^{i\vec{k} \cdot \vec{x}}
\; ,
\end{equation}
Because the self-energy conserves helicity, we may as well also specialize to
2-component helicity eigenstates,
\begin{equation}
\chi_i(\eta) = \left(\matrix{\chi_L(\eta) \cr \chi_R(\eta)} \right) \qquad
{\rm with} \qquad \vec{k} \cdot \vec{\sigma} \, \chi_{L,R} = \pm k \, 
\chi_{L,R} \; .
\end{equation}
For the left-handed spinor the result of performing the spatial integrations
is,
\begin{eqnarray}
\lefteqn{0 = (i\partial_0 \pm k) \chi_L(\eta) } \nonumber \\
& & + \frac{i \Gamma^2 H^2}{8 \pi^2} a \int_{\eta_i}^{\eta} \!\! d\eta' a' 
\chi_L(\eta') e^{\mp i k {\Delta \eta}} \left[ 2 \ln(2 H {\Delta \eta}) \!+\!
1 \!+\! \int_0^{2 k {\Delta \eta}} \!\!\!\! d\tau \frac{e^{\pm i \tau} 
\!-\! 1}{\tau} \right] . \label{nonlocal}
\end{eqnarray}
To obtain the result for a right-handed spinor with $\pm$ helicity, simply 
change the signs of $\pm k$, $\mp i k {\Delta \eta}$ and $\pm i \tau$.

Recovering the full time evolution of this sort of nonlocal mode solution
requires numerical integration. However, it is straightforward to get the
asymptotic behavior for late times. In this regime the physical 3-momentum 
has redshifted to essentially zero, so we can simplify (\ref{nonlocal}) by
setting $k=0$. This obliterates the distinctions between left and right
handedness, and between different helicities. The equation can be further 
simplified by transforming from conformal time $\eta$ to co-moving time $t$,
\begin{equation}
\partial_t \chi + \frac{\Gamma^2 H^2}{4 \pi^2} \int_0^t dt' \chi \Bigl\{
\ln\Bigl[2 e^{-H t'} - 2 e^{-H t} \Bigr] + \frac12 \Bigr\} \approx 0 \; .
\end{equation}
Now factor the leading exponential out of the logarithm,
\begin{equation}
\ln\Bigl[2 e^{-H t'} - 2 e^{-H t} \Bigr] = - H t' + \ln(2) +
\ln\Bigl[1 - e^{-H {\Delta t}}\Bigr] \; .
\end{equation}
It is an excellent approximation to retain only the term $-Ht'$. Acting 
another derivative results in a local equation,
\begin{equation}
\partial_t^2 \chi - \frac{\Gamma^2 H^2}{4 \pi^2} H t \, \chi \; , 
\end{equation}
whose approximate solution can be obtained by the WKB method,
\begin{equation}
\chi \longrightarrow (Ht)^{-\frac34} \exp\Bigl[\frac{\Gamma H}{3 \pi} 
(Ht)^{\frac32}\Bigr] \; .
\end{equation}

\end{document}